\documentclass[twoside]{article}
\usepackage[accepted]{aistatsALTERED}

\usepackage{amssymb,amsmath,natbib,dsfont,graphicx}

\newcommand{\bs}[1]{\boldsymbol{#1}}

\newcommand{\mr}[1]{\mathrm{#1}}
\newcommand{\bm}[1]{\mathbf{#1}}
\newcommand{\ds}[1]{\mathds{#1}}
\newcommand{\indep}{\perp\!\!\!\perp}

\newcommand{\cd}{\small \sf}

\begin{document}
\runningtitle{Topic Estimation and Selection}
\runningauthor{M. A. Taddy}

\twocolumn[

\aistatstitle{On Estimation and Selection for Topic Models}

\aistatsauthor{ Matthew A. Taddy }

\aistatsaddress{ Booth School of  Business\\University of Chicago \\{\cd taddy@chicagobooth.edu} }
]

\begin{abstract}
  This article describes posterior maximization for
  topic models, identifying computational and
  conceptual gains from inference under a non-standard 
  parametrization.  We then show that fitted parameters can be used
  as the basis for a novel approach to marginal likelihood estimation,
  via block-diagonal approximation to the information matrix,
  that facilitates choosing the number of latent topics.  This
  likelihood-based model selection is complemented with a
  goodness-of-fit analysis built around estimated residual dispersion.
  Examples are provided to illustrate model selection as well as to
  compare our estimation against standard alternative techniques.
\end{abstract}

\section{Introduction}
\label{introduction}

A topic model represents multivariate count data as multinomial
observations parameterized by a weighted sum of
latent {\it topics}.  With each observation $\bm{x}_i \in
\{\bm{x}_{1}\ldots \bm{x}_n\}$ a vector of counts in $p$ categories,
given total count $m_i = \sum_{j=1}^p x_{ij}$,
the  $K$-topic model has
\begin{equation}\label{eq:tpc}
\bm{x}_i \sim \mr{MN}(\omega_{i1} \bs{\theta}_{1} + \ldots + \omega_{iK}
\bs{\theta}_{K}, m_i)
\end{equation}
where topics $\bs{\theta}_k = [\theta_{k1} \cdots \theta_{kp}]'$ and
weights $\bs{\omega}_i$ are probability vectors.  The {\it topic}
label is due to application of the model in (\ref{eq:tpc}) to the
field of text analysis.  In this context, each $\bm{x}_i$ is a vector
of counts for terms (words or phrases) in a {\it document} with total
term-count $m_i$, and each topic $\bs{\theta}_k$ is a vector of
probabilities over words.  Documents are thus characterized through a
mixed-membership weighting of topic factors and, with $K$ far smaller
than $p$, each $\bs{\omega}_i$ is a reduced dimension summary for
$\bm{x}_i$.

Section \ref{background} surveys the wide use of topic models and
emphasizes some aspects of current technology that show room for
improvement.  Section \ref{background}.1 describes how common
large-data estimation for topics is based on maximization of an
approximation to the marginal likelihood, $\mr{p}(\bm{X}|
\bs{\Theta})$.  This involves very high-dimensional latent variable
augmentation, which complicates and raises the cost of computation,
and independence assumptions in approximation that can potentially
bias estimation.  Moreover, Section \ref{background}.2 reviews the
very limited literature on topic selection, arguing the need of new
methodology for choosing $K$.

The two major objectives of this article are thus to facilitate an
efficient alternative for estimation of topic models, and to provide a
default method for model selection.  In the first case, Section
\ref{estimation} develops a framework for joint posterior maximization
over both topics and weights: given the parameterization described in
\ref{estimation}.1, we outline a block-relaxation algorithm in
\ref{estimation}.2 that augments expectation-maximization with
quadratic programming for each $\bs{\omega}_i$.  Section
\ref{modelchoice} then presents two possible metrics for model choice:
\ref{modelchoice}.1 describes marginal data likelihood estimation
through block-diagonal approximation to the information matrix, while
\ref{modelchoice}.2 proposes estimation for residual dispersion.  We
provide simulation and data examples in Section 5 to support and
illustrate our methods, and close with a short discussion in Section
\ref{discussion}.

\section{Background}
\label{background}

The original text-motivated topic model is due to \citet{Hofm1999},
who describes its mixed-membership likelihood as a probability model
for the latent semantic indexing of \citet{DeerDumaFurnLandHars1990}.
\citet{BleiNgJord2003} then introduce the contemporary Bayesian
formulation of topic models as latent Dirichlet allocation (LDA) by
adding conditionally conjugate Dirichlet priors for topics and
weights.  This basic model has proven hugely popular, and extensions
include hierarchical formulations to account for an unknown number of
topics \citep[][using Dirichlet processes]{TehJordBealBlei2006},
topics that change in time \citep{BleiLaff2006} or whose expression is
correlated \citep{BleiLaff2007}, and topics driven by sentiment
\citep{BleiMcAu2010}.  Equivalent likelihood models, under both
classical and Bayesian formulation, have also been independently
developed in genetics for analysis of population admixtures
\citep[e.g.,][]{PritStepDonn2000}.

\subsection{Estimation Techniques}

This article, as in most text-mining applications, 
focuses on a Bayesian specification of the model in (\ref{eq:tpc}) with
independent priors for each $\bs{\theta}_k$ and $\bs{\omega}_i$,
yielding a posterior distribution proportional to
\begin{equation}\label{eq:joint} 
\mr{p}(\bs{\Theta},\bs{\Omega},\bm{X}) =
\prod_{i=1}^n \mr{MN}(\bm{x}_i ; \bs{\Theta}\bs{\omega}_i, m_i)
\mr{p}(\bs{\omega}_i) \prod_{k=1}^K \mr{p}(\bs{\theta}_k).
\end{equation}
Posterior approximations rely on augmentation with topic
membership for individual terms: assume term $l$ from document $i$ has
been drawn with probability given by topic $\bs{\theta}_{z_{il}}$,
where membership $z_{il}$ is sampled from the $K$ available according
to $\bs{\omega}_i$, and write $\bm{z}_i$ as the $m_i$-length indicator
vector for each document $i$ and $\bm{Z} = \{ \bm{z}_1,
\ldots,\bm{z}_n\}$ as the full latent parameter matrix.

Gibbs sampling for this specification is described in the original
genetic admixture paper by \citet{PritStepDonn2000} and by some in
machine learning \citep[e.g.,][]{GrifStyv2004}.  Many software
packages adapt Gibbs to large-data settings by using a single
posterior draw of $\bm{Z}$ for parameter estimation.  This only
requires running the Markov chain until it has reached its stationary
distribution, instead of the full convergence assumed in mean
estimation, but the estimates are not optimal or consistent in any
rigorous sense.

It is more common in practice to see variational posterior
approximations \citep{WainJord2008}, wherein some tractable
distribution is fit to minimize its distance from the unknown true
posterior.  For example, consider the conditional posterior
$\mr{p}(\bs{\Omega}, \bm{Z} \mid \bs{\Theta}, \bm{X})$ and variational
distribution $\mr{q}( \bs{\Omega}, \bm{Z} ) = \prod_{i=1}^n\left[
  \mr{Dir}(\bs{\omega_i}; \bm{s}_i) \prod_{l=1}^{m_i} \mr{MN}(z_{il};
  \bm{r}_{il}) \right]$.  Kullback-Leibler divergence between these
densities is
\begin{align}\label{eq:kl}
-&\int \log\frac{\mr{p}(\bs{\Omega}, \bm{Z}
    ~|~\bs{\Theta},\bm{X})}{\mr{q}(\bs{\Omega}, \bm{Z})}
d\mr{Q}(\bs{\Omega},\bm{Z})\\
&~~=  \log\mr{p}(\bm{X}\mid \bs{\Theta}) \notag \\ &~~~~~~~ - \left\{\ds{E}_q\left[ \log
    \mr{p}(\bs{\Omega}, \bm{Z}, \bm{X}\mid \bs{\Theta}) \right] - \ds{E}_q\left[
  \log \mr{q}(\bs{\Omega},\bm{Z}) \right] \right\}, \notag
\end{align}
which is minimized by maximizing the lower bound on marginal
likelihood for $\bs{\Theta}$, $\ds{E}_q\left[ \log \mr{p}(\bs{\Omega},
  \bm{Z}, \bm{X}\mid \bs{\Theta}) \right] - \ds{E}_q\left[ \log
  \mr{q}(\bs{\Omega},\bm{Z}) \right] $, as a function of $\mr{q}$'s
tuning parameters.  Since $\bs{\omega}_i \indep \bs{\omega}_l$ for
$i\neq l$ conditional on $\bs{\Theta}$, $\mr{q}$'s main relaxation is
assumed independence of $\bm{Z}$.

The objective in (\ref{eq:kl}) is proposed in the original LDA paper
by \cite{BleiNgJord2003}.\footnote{\citet{TehNewmWell2006} describe
  alternative approximation for the marginal indicator posterior,
  $\mr{p}(\bm{Z} \mid \bm{X})$; this avoids conditioning on
  $\bs{\Theta}$, but keeps independence assumptions over $\bm{Z}$ that
  will be even less accurate than they are in the conditional
  posterior.}  The most common approach to estimation
\citep[e.g.,][and examples in
\citealt{BleiNgJord2003}]{BleiLaff2007,Grim2010} is then to maximize
$\ds{E}_q\left[ \log \mr{p}(\bs{\Omega}, \bm{Z}, \bm{X} \mid
  \bs{\Theta}) \right]$ given $\mr{q}( \bs{\Omega}, \bm{Z} )\approx
\mr{p}(\bs{\Omega}, \bm{Z} ~|~\bs{\Theta},\bm{X})$.  This {\it
  mean-field estimation} can be motivated as fitting
$\bs{\widehat\Theta}$ to maximize the implied lower bound on
$\mr{p}(\bm{X} \mid \bs{\Theta})$ from (\ref{eq:kl}), and thus
provides approximate marginal maximum likelihood estimation.  The
procedure is customarily identified with its implementation as a
variational EM (VEM) algorithm, which iteratively alternates between
conditional minimization of (\ref{eq:kl}) and maximization of the
bound on $\mr{p}(\bm{X} ~|~\bs{\Theta})$.

A second strategy,
full variational Bayes, constructs a joint distribution
through multiplication of $\mr{q}(\bs{\Omega}, \bm{Z})$ by
$\mr{q}(\bs{\Theta}) = \prod_{k=1}^K \mr{Dir}(\bs{\theta}_k \mid
\bm{u}_k)$ and minimizes KL
divergence against the full posterior.  However, since cross-topic
posterior correlation $\mr{cor}(\theta_{kj},\theta_{hj})$ does not
disappear asymptotically\footnote{See the information matrix in
\ref{modelchoice}.1.}, the independence assumption of
$\mr{q}(\bs{\Theta})$ risks inconsistency for $\bs{\widehat\Theta}$
and unstable finite sample results \citep[e.g.,][]{TehNewmWell2006}. 

Our proposed approach is distinct from the above in seeking joint MAP
solution for both $\bs{\Theta}$ and $\bs{\Omega}$ (i.e., that which
maximizes (\ref{eq:joint})), thus altogether avoiding posterior
approximation.  The estimation methodology of Section \ref{estimation}
is more closely connected to maximum likelihood estimation (MLE)
techniques from the non-Bayesian literature on topic models: the EM
algorithm, as used extensively for genetics admixture estimation
\citep[e.g.,][]{TangCoraWangZhuRisc2006} and in Hoffman's 1999
text-analysis work, and quadratic programming as applied by
\cite{AlexNoveLang2009} in a fast block-relaxation routine.  We borrow
from both strategies.

Comparison between MAP and VEM estimation is simple: the former finds
jointly optimal profile estimates for $\bs{\Theta}$ and $\bs{\Omega}$,
while the latter yields inference for $\bs{\Theta}$ that is
approximately integrated over uncertainty about $\bs{\Omega}$ and
$\bm{Z}$.  There are clear advantages to integrating over nuisance
parameters \citep[e.g.,][]{BergLiseWolp1999}, but marginalization
comes at the expense of introducing a very high dimensional latent
parameter ($\bm{Z}$) and its posterior approximation.  This leads to
algorithms that are not scaleable in document length, potentially with
higher variance estimation or unknown bias.  We will find that, given
care in parameterization, exact joint parameter estimation can be
superior to approximate marginal inference.

\subsection{Choosing the Number of Topics}

To learn the number of topics from data, the literature makes use of
three general tools: cross-validation, nonparametric mixture priors,
and marginal likelihood.  See \cite{AiroErosFienJout2010} for a short
survey and comparison.  Cross-validation (CV) is by far the most
common choice \citep[e.g.][]{Grim2010,GrunHorn2011}.  Unfortunately,
due to the repeated model-fitting required for out-of-sample
prediction, CV is not a large-scale method.  It also lacks easy
interpretability in terms of statistical evidence, or even in terms of
sample error \citep[][7.12]{HastTibsFrie2009}.  For a model-based
alternative, \cite{TehJordBealBlei2006} write LDA as a Hierarchical
Dirichlet process with each document's weighting over topics a
probability vector of infinite length.  This removes the need to
choose $K$, but estimation can be sensitive to the level of finite
truncation for these prior processes and will always require inference
about high-dimensional term-topic memberships.  Finally, the standard
Bayesian solution is to maximize the marginal model posterior.  However,
marginal likelihood estimation in topic models has thus far been
limited to very rough approximation, such as the average of MCMC draws
in \citep[][]{GrifStyv2004}.

\section{Parameter Estimation}
\label{estimation}

Prior choice for latent topics can be contentious
\citep{WallMimnMcCa2009} and concrete guidance is lacking in the
literature.  We consider simple independent priors, but estimation updates based
on conditional independence make it straightforward to adapt for more
complex schemes.  Our default specification follows from a general
preference for simplicity.  Given $K$ topics,
\begin{align}\label{eq:prior}
\bs{\omega}_i &\stackrel{iid}{\sim} \mr{Dir}(1/K),~i=1\ldots n,\\
\bs{\theta}_k  &\stackrel{iid}{\sim}
\mr{Dir}(\alpha_{k1}, \ldots, \alpha_{kp}),~k=1\ldots K.\notag
\end{align}
The single Dirichlet concentration parameter of $1/K$ for each
$\bs{\omega}$ encourages sparsity in document weights by placing prior
density at the edges of the parameter space.  This specification is
also appropriate for model selection: weight of prior evidence is
constant for all values of $K$, and as $K$ moves to infinity
$\mr{Dir}(1/K)$ approaches the Dirichlet process \citep{Neal2000}.
Topic priors are left generic in (\ref{eq:prior}) to encourage
flexibility
but we default to the low concentration $\alpha_{kj} = 1/(Kp)$.
 
\subsection{Natural Exponential Family Parameterization}

To improve estimation stability and efficiency, we propose to solve
for MAP estimates of $\bs{\Omega}$ and $\bs{\Theta}$ not in the
original simplex space, but rather after transform into their natural
exponential family (NEF) parameterization.  For example, in the case
of $\bs{\Omega}$ we seek the MAP estimate for $\bs{\Phi} = \{
\bs{\varphi}_1, \ldots, \bs{\varphi}_n\}$, where for a given
$\bs{\omega}$,
\begin{equation}\label{eq:nef}
  \omega_k = \frac{\exp[{\varphi_{k-1}}]}{\sum_{h=0}^{K-1}
    \exp[{\varphi_{h}}]},
~~\text{with~the~fixed~element}~~\varphi_0=0.
\end{equation}
Hence, ignoring $\varphi_{i0}$ in estimation, each $\bs{\varphi}_i $
is an unrestricted vector of length $K-1$.  Since $\partial \omega_k
/ \partial \varphi_h = \ds{1}_{k=h}\omega_k - \omega_k \omega_h$, the
Jacobian for this transformation is $\mr{Diag}[\bs{\omega}] -
\bs{\omega}\bs{\omega}'$ and has determinant
$|\mr{Diag}[\bs{\omega}]|(1 -\bs{\omega}'
\mr{Diag}[\bs{\omega}]\bs{\omega}) =
\prod_{k=1}^{K-1}\omega_k(1-\sum_{h=1}^{K-1}\omega_h)$.  Hence,
viewing each $\bs{\omega}_i$ as a function of $\bs{\varphi}_i$, the
conditional posterior for each individual document given $\bs{\Theta}$
becomes
\begin{align}\label{eq:nefpost}
\mr{p}&(\bs{\omega}_i(\bs{\varphi}_i) \mid \bm{x}_i) \\ &\propto
\mr{MN}(\bm{x}_i ; \bs{\Theta}\bs{\omega}_i, m_i)
\prod_{k=1}^{K-1}\omega_k^{1/K} \left(1-\sum_{h=1}^{K-1}\omega_h\right)^{1/K},\notag
\end{align}
and the NEF conditional MAP is equivalent to solution for
$\bs{\omega}$ under a $\mr{Dir}(1/K + 1)$ prior.  Similarly, our
estimate for each $\bs{\theta}_k$ corresponds to the simplex MAP under
a $\mr{Dir}(\bs{\alpha}_k+1)$ prior.\footnote{NEF parameterization
  thus removes the ``-1 offset'' for MAP estimation critiqued by
  \citet{AsunWellSmytTeh2009} wherein they note that different
  topic model algorithms can be made to provide  similar fits through
  prior-tuning.}

The NEF transformation leads to conditional posterior functions that
are everywhere concave, thus guaranteeing a single conditional MAP
solution for each $\bs{\omega}_i$ given $\bs{\Theta}$.  This
introduces stability into our block relaxation algorithm of
\ref{estimation}.2: without moving to NEF space, the prior in
(\ref{eq:prior}) with $1/K < 1$ could lead to ill-defined maximization
problems at each iteration.  Conditional posterior concavity also
implies non-boundary estimates for $\bs{\Omega}$, despite our use of
sparsity encouraging priors, that facilitate Laplace approximation in
Section \ref{modelchoice}.1.

\subsection{Joint Posterior Maximization}

We now detail joint MAP estimation for $\bs{\Omega}$ and $\bs{\Theta}$
under NEF parameterization.  First, note that it is
straightforward to build an EM algorithm around missing data
arguments.  In the MLE literature
\citep[e.g.,][]{Hofm1999,TangCoraWangZhuRisc2006} authors use the full
set of latent phrase-memberships ($\bm{Z}$) to obtain a mixture model
specification.  However, a lower dimensional strategy is based on only
latent topic {\it totals}, such that each document $i = 1\ldots n$ is
expanded
\begin{equation}
  \bm{X}_i \sim  \mr{MN}_p(\bs{\theta}_1, t_{i1}) + \cdots +
  \mr{MN}_p(\bs{\theta}_K, t_{iK}),
\end{equation}
where $\bm{T}_i\sim \mr{MN}_K(\bs{\omega}_i, m_i)$, with
$\bm{T}_1\ldots\bm{T}_n$ treated as missing-data.  Given current
estimates $\bs{\widehat{\Theta}}$ and $\bs{\widehat{\Omega}}$,
standard EM calculations lead to approximate likelihood bounds of
$\mr{MN}\left( \bm{\hat X}_k; \bs{\theta}_k, \hat{t}_k \right) $, with
\begin{equation}\hat{x}_{kj} = \sum_{i=1}^n
x_{ij}\frac{\hat\theta_{kj}\hat\omega_{ik}} {\sum_{h=1}^K
  \hat\theta_{hj}\hat\omega_{ih}},~~ \hat{t}_k = \sum_{j=1}^p
  \hat{x}_{kj},~~k=1\ldots K. \label{EM}
\end{equation}
Updates for NEF topic MAPs under our model are then $ \theta_{kj} =
(\hat{x}_{kj}+\alpha_{kj})/[\hat{t}_k+ \sum_{j=1}^p\alpha_{kj}]$.

EM algorithms -- even the low dimensional version in (\ref{EM}) -- are
slow to converge under large-$p$ vocabularies.  We advocate finding
exact solutions for $\bs{\Omega} \mid \bs{\Theta}$ at each iteration,
as this can speed-up convergence by several orders of
magnitude.  Factorization of the conditional posterior makes for fast
parallel updates that, similar to the algorithm of
\citet{AlexNoveLang2009}, solve independently for each $\bs{\omega}_i$
through sequential quadratic programming.\footnote{Alexander et
  al. also use this approach to update $\bs{\Theta} \mid \bs{\Omega}$
  in a similar model, but we have found in our applications that any
  advantage over EM for $\bs{\Theta}$ is out-weighed by computational
  expense in the high-dimensional pivoting required by constraints
  $\sum_{j=1}^p \theta_{kj} = 1$.}  We also add first-order
quasi-Newton acceleration for full-set updates \citep{Lang2010}.

Suppressing $i$, each document's conditional log posterior is
proportional to
\begin{equation}\label{logpost}
l(\bs{\omega}) =  \sum_{j=1}^p x_{j}
  \log\left(\omega_{1}\theta_{1j} + \ldots +\omega_{K}\theta_{Kj}
  \right) + \sum_{k=1}^K \frac{\log(\omega_k)}{K},
\end{equation}
subject to the constraints $\bs{1}'\bs\omega = 1$ and $\omega_k > 0$
for $k=1\ldots K$.  Gradient and
curvature are then
\begin{align}\label{dW}
g_k &= \sum_{j=1}^p \frac{x_j\theta_{kj}}{\bs{\theta}'_j\bs{\omega}} +\frac{1}{K\omega_k}\\
h_{kh} &= -\sum_{j=1}^p
\frac{x_j\theta_{kj}\theta_{hj}}{(\bs{\theta}'_j\bs{\omega})^2} \notag
- \ds{1}_{[k=h]}\frac{1}{K\omega_k^2}
\end{align}
and Taylor approximation around current estimate $\bs{\hat\omega}$ yields the linear
system
\begin{equation}\label{system}
\left[\begin{array}{cc}
-\bm{h} & \bs{1} \\
\bs{1}' & 0
  \end{array} \right]
  \left[
\begin{array}{c} 
\bs{\Delta}\\
\lambda
  \end{array} \right] 
= \left[\begin{array}{c} \bm{g} \\ 0 
  \end{array} \right]
\end{equation}
where $\Delta_k = (\omega_k - \hat\omega_k)$ and $\lambda$ is the
Lagrange multiplier for the equality constraint.  This provides the
basis for an active set strategy \citep[][12.3]{Luen2008}: given
$\bs{\Delta}$ solving (\ref{system}), take maximum $\delta \in (0,1)$
such that $\delta\Delta_k < - \hat\omega_k~\forall k$, and set
$\bs{\omega} = \bs{\hat\omega} + \delta\bs{\Delta}$.  If $\delta < 1$
and $\omega_k$ lies at boundary of the feasible region (i.e., 
some tolerance from zero), we activate that constraint by removing
$\Delta_k$ and solving the simpler system.

Note from (\ref{dW}) that our log posterior in (\ref{logpost}) is
concave, thus guaranteeing a unique solution at every 
iteration.  This would not be true but for the fact that we are
actually solving for conditional MAP $\bs{\varphi}$, rather than
$\bs{\omega}$.  While the full joint posterior for transformed
$\bs{\Theta}$ and $\bs{\Omega}$ obviously remains multi-modal (most
authors use multiple starts to avoid minor modes; we initialize with a
build from $2,\ldots, K$ topics by repeatedly fitting an extra topic
to the residuals), the NEF parameterization introduces helpful
stability at each iteration.

\section{Model Selection}
\label{modelchoice}

In this section, we propose two techniques for inferring the number of
topics.  The first approach, in \ref{modelchoice}.1, is an efficient
approximation for the fully Bayesian procedure of marginal posterior
maximization.  The second approach, in \ref{modelchoice}.2, consists
of a basic analysis of residuals.  Both methods require almost no
computation beyond the parameter estimation of Section
\ref{estimation}.

\subsection{Marginal Likelihood Maximization}

Bayesian model selection is founded on the marginal data likelihood
\citep[see][]{KassRaft1995}, and in the absence of a null hypothesis
scenario or an informative model prior, we wish to find $K$ to
maximize $\mr{p}( \bm{X} \mid K) = \int\mr{p}(\bs{\Theta},\bs{\Omega},
\bm{X} \mid K)d\mr{P}(\bs{\Theta},\bs{\Omega})$.  It should be
possible to find a maximizing argument simply by evaluating
$\mr{p}(\bm{X}\mid K)$ over possible values.  However, as is
often the case, the integral is intractable for topic
models and must be approximated.  

One powerful approach is Laplace's method \citep{TierKada1986}
wherein, assuming that the posterior is highly peaked around its
 mode, the joint parameter-data likelihood is replaced with a
close-matching and easily integrated Gaussian density.  In particular,
quadratic expansion of $\log[\mr{p}(\bm{X}, \bs{\Phi}, \bs{\Theta})]$
is exponentiated to yield the approximate posterior,
$\mr{N}([\bs{\Phi}, \bs{\Theta}]; [\bs{\hat\Phi}, \bs{\hat\Theta}],
\bm{H})$, where $[\bs{\hat\Phi}, \bs{\hat\Theta}]$ is the joint MAP
and $\bm{H}$ is the log posterior Hessian evaluated at this
point.  After scaling by $K!$ to account for label switching, these
approximations have proven effective 
for general mixtures \cite[e.g.][]{RoedWass1997}.
 
As one of many advantages of NEF parameterization
\citep[][]{Mack1998}, we avoid boundary solutions where Laplace's
approximation would be invalid.  The integral target is also lower
dimensional, although we leave $\bs{\Theta}$ in simplex representation
due to a denser and harder to approximate Hessian after NEF transform.
Hence, 
\begin{equation}\label{laplace}
\mr{p}(\bm{X} \mid K) \approx \mr{p}\left(\bm{X},\bs{\hat \Theta}, \bs{\hat
  \Omega}\right)|-\bm{H}|^{-\frac{1}{2}}(2\pi)^{\frac{d}{2}} K!
\end{equation}
where $d = Kp + (K-1)n$ is
model dimension and $\mr{p}(\bm{X},\bs{\hat \Theta}, \bs{\hat \Omega}) 
=
\prod_{i=1}^n \mr{MN}( \bm{x}_i ; 
\bs{\hat\Theta}\bs{\hat\omega}_i, m_i) 
\mr{Dir}(\bs{\hat\omega}_i; 1/K\!+\!1)  $ $
\prod_{k=1}^K
\mr{Dir}(\bs{\hat\theta}_k; \bs{\alpha}_k+1)$.  In practice, to account for weight
sparsity, we replace $d$ with $Kp + d_{\bs{\Omega}}$ where
$d_{\bs{\Omega}}$ is the number of $\omega_{ik}$ greater than $1/1000$.

After parameter estimation, the only additional computational burden
of (\ref{laplace}) is finding the determinant of negative $\bm{H}$.
Unfortunately, given the large $p$ and $n$ of text analysis, this
burden will usually be significant and unaffordable.  Determinant
approximation for marginal likelihoods is not uncommon; for
example, {\it Bayes information criterion} (BIC) can be motivated by
setting $|-\bm{H}| \approx n^d |\bm{i}|$, with $\bm{i}$ the
information matrix for a single observation. Although BIC convergence
results do not apply under $d$ that depends on $n$, efficient
computation is possible through a more subtle approximation based on
block-diagonal factorization.

Writing $L = \log \left[ \mr{p}(\bm{X},\bs{ \Theta}, \bs{\Omega})
\right]$, diagonal blocks of $\bm{H}$ are
\begin{align} \label{diag}
&\bm{H}_{\bs{\Theta}} = \frac{\partial^2L}{\partial
\bs{\Theta}^2} =
\mr{Diag}\left[
 \displaystyle \frac{\partial^2L}{\partial \bs{\theta}_{\bullet
     1}^2} \cdots
\displaystyle \frac{\partial^2L}{\partial \bs{\theta}_{\bullet  p}^2}
\right]\notag\\
&\bm{H}_{\bs{\Phi}} = \frac{\partial^2L}{\partial
\bs{\Phi}^2} =
\mr{Diag}\left[
 \displaystyle \frac{\partial^2L}{\partial \bs{\varphi}_{
     1}^2}\cdots \frac{\partial^2L}{\partial \bs{\varphi}_{n}^2}
\right]
\end{align}
where $\bs{\theta}_{\bullet j} = [ \theta_{1j} \cdots \theta_{Kj} ]$
is the $j^{th}$ row of $\bs{\Theta}$. Here,  
\[
\frac{\partial^2L}{\partial
\theta_{kj}\partial \theta_{hj}} = \sum_{i=1}^n x_{ij}
\frac{\omega_{ik}\omega_{ih}}{q_{ij}^2} + \ds{1}_{k=h}
\frac{\alpha_{jk}}{\theta_{kj}^2},
\]
while for each individual document's $\bs{\varphi}_i$,
\begin{align*}&
\frac{\partial^2L}{\partial
\varphi_{ik}\partial \varphi_{ih}} = \ds{1}_{[k=h]}\omega_k
- \omega_{ik}\omega_{ih}\\
& -\sum_{j=1}^p x_{ij}
\left[ \ds{1}_{[k=h]} \omega_{ik} \frac{\theta_{kj} - q_{ij}}{q_{ij}} +  
\omega_{ik}\omega_{ih}\left(1 -
  \frac{\theta_{kj}\theta_{hj}}{q_{ij}^2}\right)\right] 
\end{align*}
with $q_{ij} = \sum_{k=1}^K \omega_{ik}\theta_{kj}$.  Finally, the
sparse off-diagonal blocks of $\bm{H}$ have elements 
\[
\frac{\partial^2 L}{\partial \theta_{jk} \partial \varphi_{ih}} = -x_{ij}\left[
  \frac{\omega_{ik}\omega_{ih}}{q_{ij}^2}\theta_{hj} -
  \frac{\omega_{ih}}{q_{ij}} \ds{1}_{[k=h]} \right]
\]
wherever $x_{ij} \neq 0$, and zero otherwise.  Ignoring these cross
curvature terms, an approximate determinant is available as the
product of determinants for each diagonal block in (\ref{diag}).  That
is, our marginal likelihood estimate is as in equation
(\ref{laplace}), but with replacement
\begin{equation}\label{approxH}
\left|-\bm{H}\right| \approx \left| 
\begin{array}{cc} -\bm{H}_{\bm{\Theta}} & \bm{0}\\
\bm{0} & -\bm{H}_{\bm{\Phi}} 
\end{array} \right| = 
\prod_{j=1}^p \left|-\frac{\partial^2L}{\partial \bs{\theta}_{\bullet
     j}^2}\right|\prod_{i=1}^n \left|-\frac{\partial^2L}{\partial \bs{\varphi}_{
     i}^2} \right|.
\end{equation}
This is fast and easy to calculate, and we show in Section
\ref{examples} that it performs well in finite sample examples.
Asymptotic results for block diagonal determinant approximations are
available in \citet{IpseLee2011}, including convergence rates and
higher-order expansions.  From the statistician's perspective, very
sparse off-diagonal blocks contain only terms related to covariance
between shared topic vectors and an individual document's weights, and
we expect that as $n \rightarrow \infty$ the influence of these
elements will disappear.

\subsection{Residuals and Dispersion}

Another strategy is to consider the simple connection
between number of topics and model fit: given theoretical
multinomial dispersion of $\sigma^2 =1$, and conditional on the
topic-model data generating process of (\ref{eq:tpc}), any fitted
overdispersion $\hat\sigma^2 > 1$ indicates a true $K$ that
is larger than the number of estimated topics.

Conditional on estimated probabilities $\bm{\hat q}_i =
\bs{\hat\Theta}\bs{\hat\omega}_i$, each document's fitted phrase
counts are $\hat x_{ij} = \hat q_{ij}m_i$.  Dispersion $\sigma^2$ can
be derived from the relationship $\ds{E}\left[ (x_{ij} -
  \hat{x}_{ij})^2\right] \approx \sigma^2 m_i
\hat{q}_{ij}(1-\hat{q}_{ij})$ and
estimated following \citet{Habe1973} as the mean of squared {\it
  adjusted residuals} $(x_{ij} - \hat{x}_{ij})/s_{ij}$, where
$s_{ij}^2 = m_i\hat{q}_{ij}(1-\hat{q}_{ij})$.  Sample dispersion
is then $\hat \sigma^2 = D/\nu$, where
\begin{equation}\label{dispersion}
D 
=  \sum_{\{i,j:~x_{ij} >0\}}
  \frac{x_{ij}^2 - 2x_{ij}\hat
    x_{ij}}{m_i\hat{q}_{ij}(1-\hat{q}_{ij})}
+ \sum_{i=1}^n\sum_{j=1}^p m_i \frac{\hat{q}_{ij}}{1-\hat{q}_{ij}}
\end{equation}
and $\nu$ is an estimate for its degrees of freedom.  We use $\nu =
\hat N - d$, with $\hat N$ the number of $\hat x_{ij}$ greater than
1/100.  $D$ also has approximate $\chi^2_\nu$ distribution under the
hypothesis that $\sigma^2 = 1$, and a test of this against alternative
$\sigma^2 >1$ provides a very rough measure for evidence in favor of a
larger number of topics.

\begin{figure*}[t!]
\vskip -.2cm
\hskip .75cm ~(a)\hskip -.6cm \includegraphics[height=2.7in]{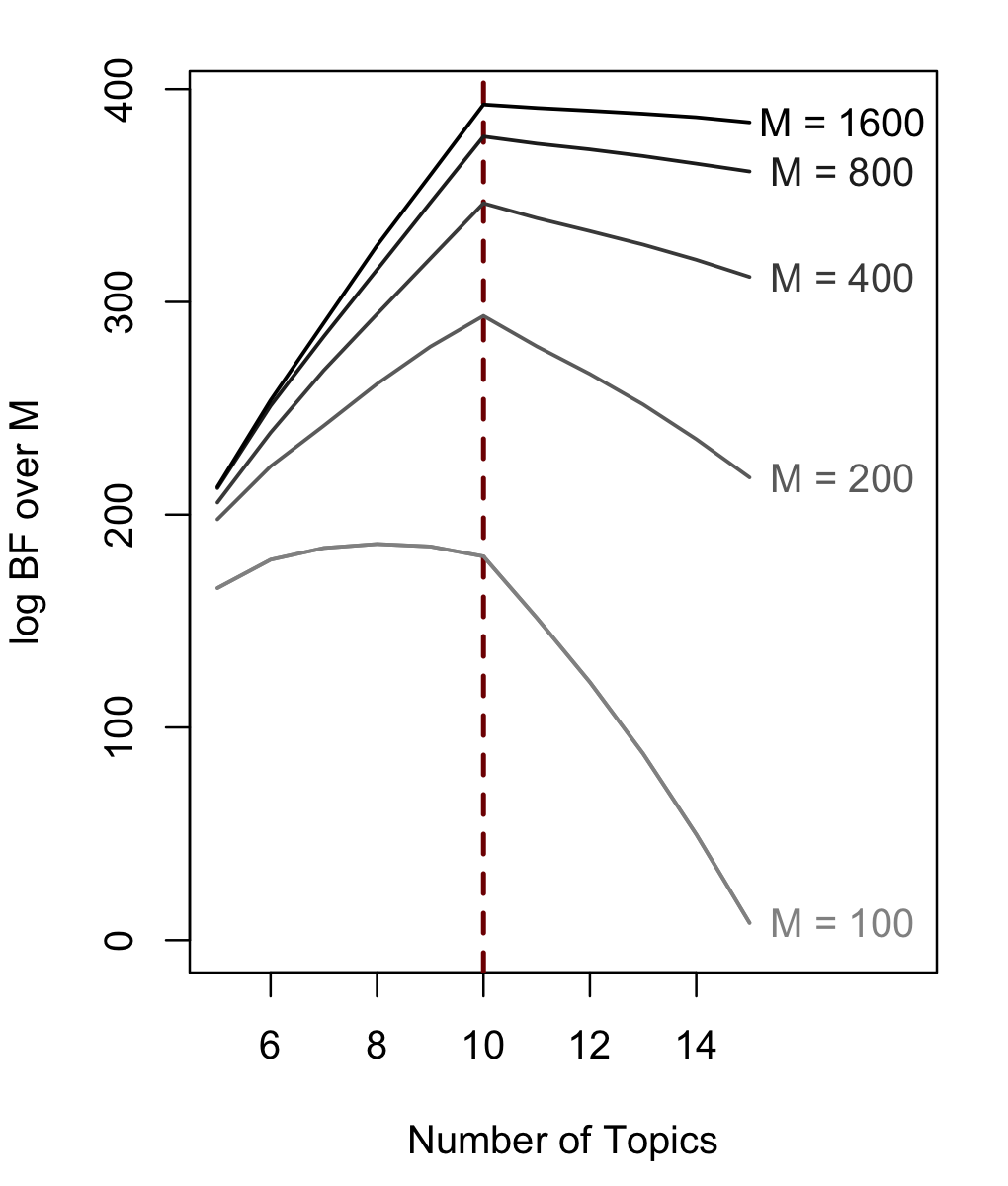}\hskip 2cm
(b)\hskip -.7cm\includegraphics[height=2.7in]{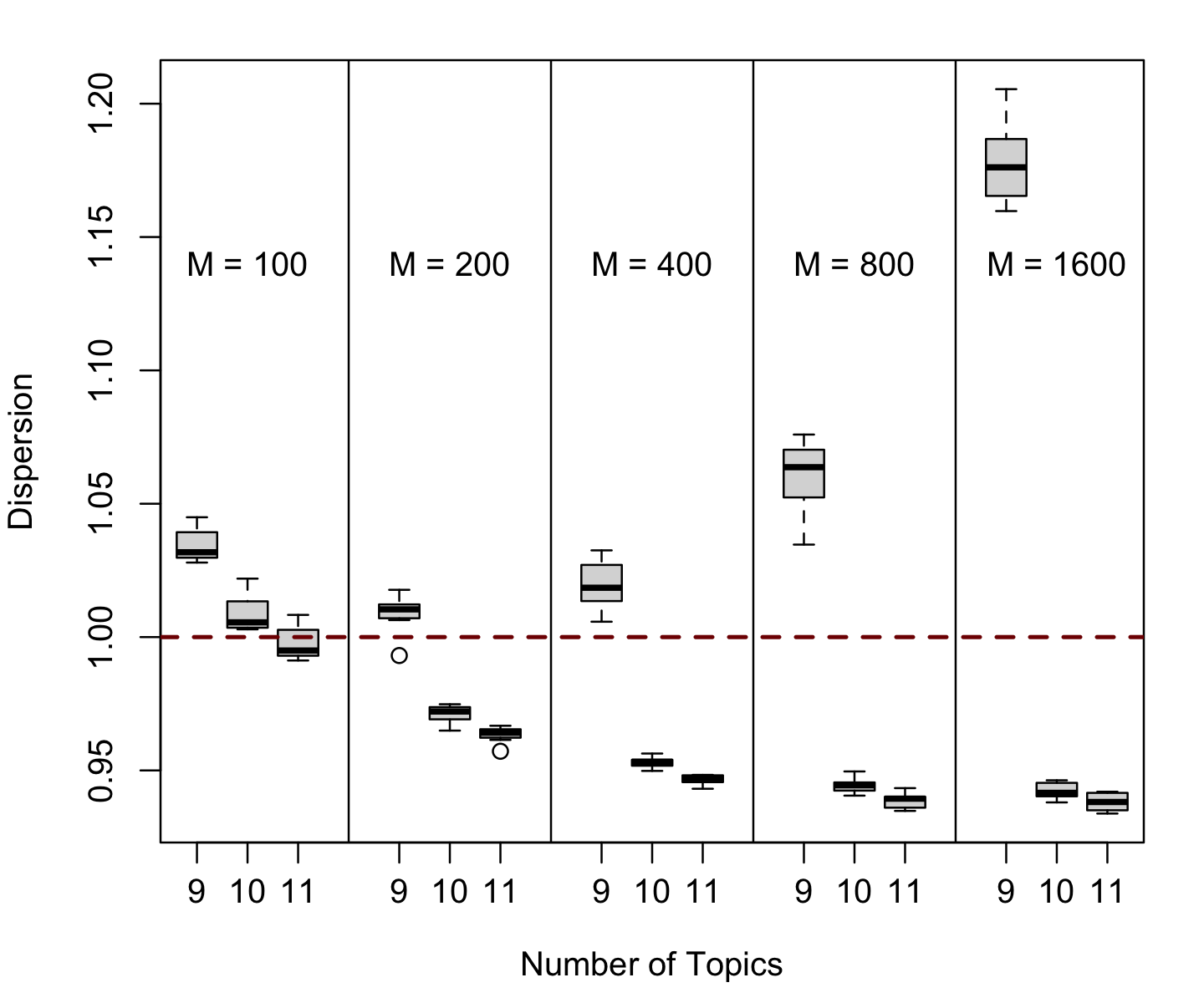}\\
\vskip -.4cm
\caption{\label{simtopics} From simulation under various values of
  $M$, the expected document size, plot (a) shows average log Bayes
  factors for $K=$ 5-15 against the null one-topic model, and (b) shows
  estimated dispersion for $K=$ 9-11. }
\vskip .1cm
\hskip 0.2cm (a)\hskip -1.1cm \includegraphics[height=2.55in]{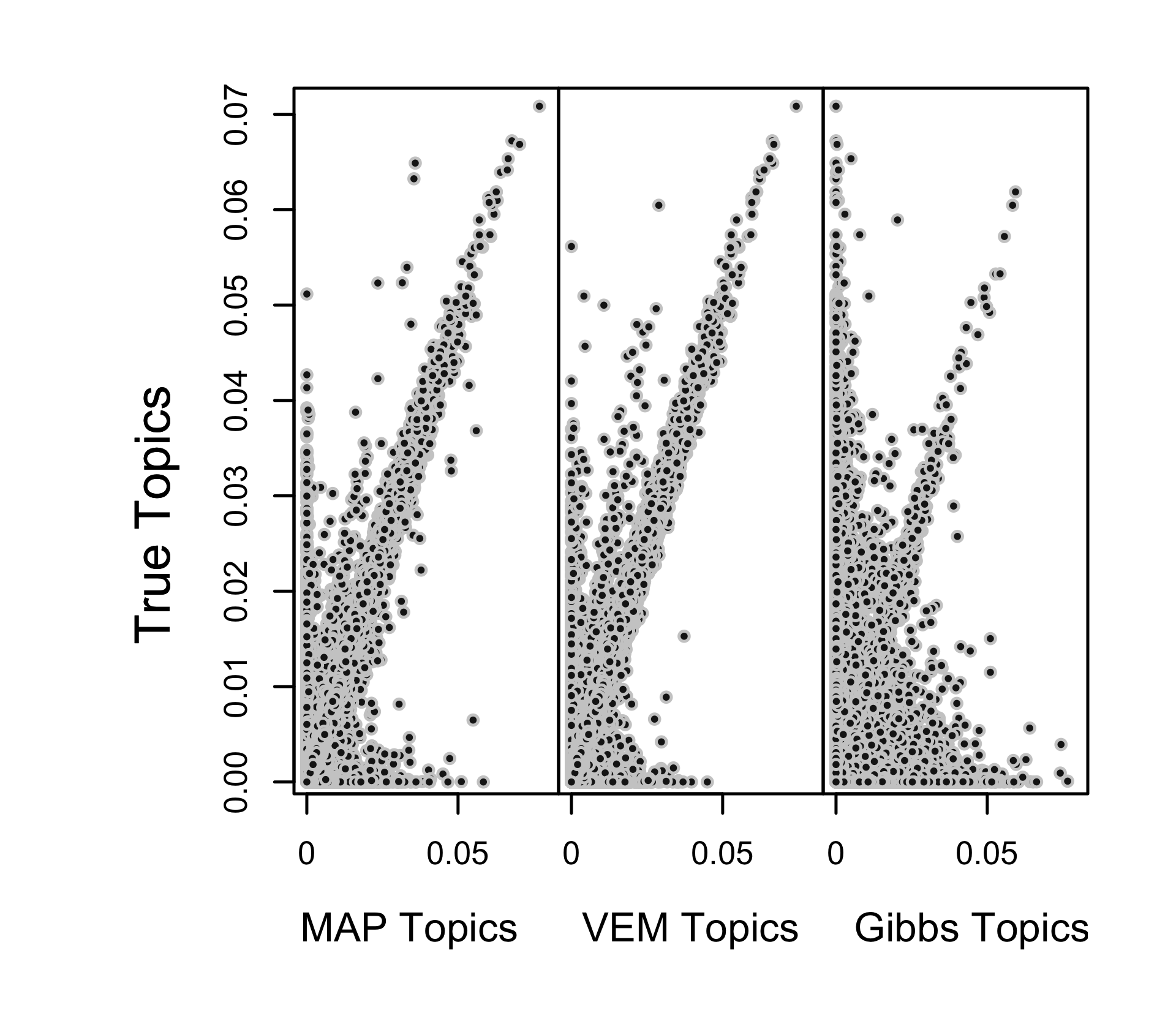}\hskip .5cm
(b)\hskip
-.4cm\includegraphics[height=2.5in]{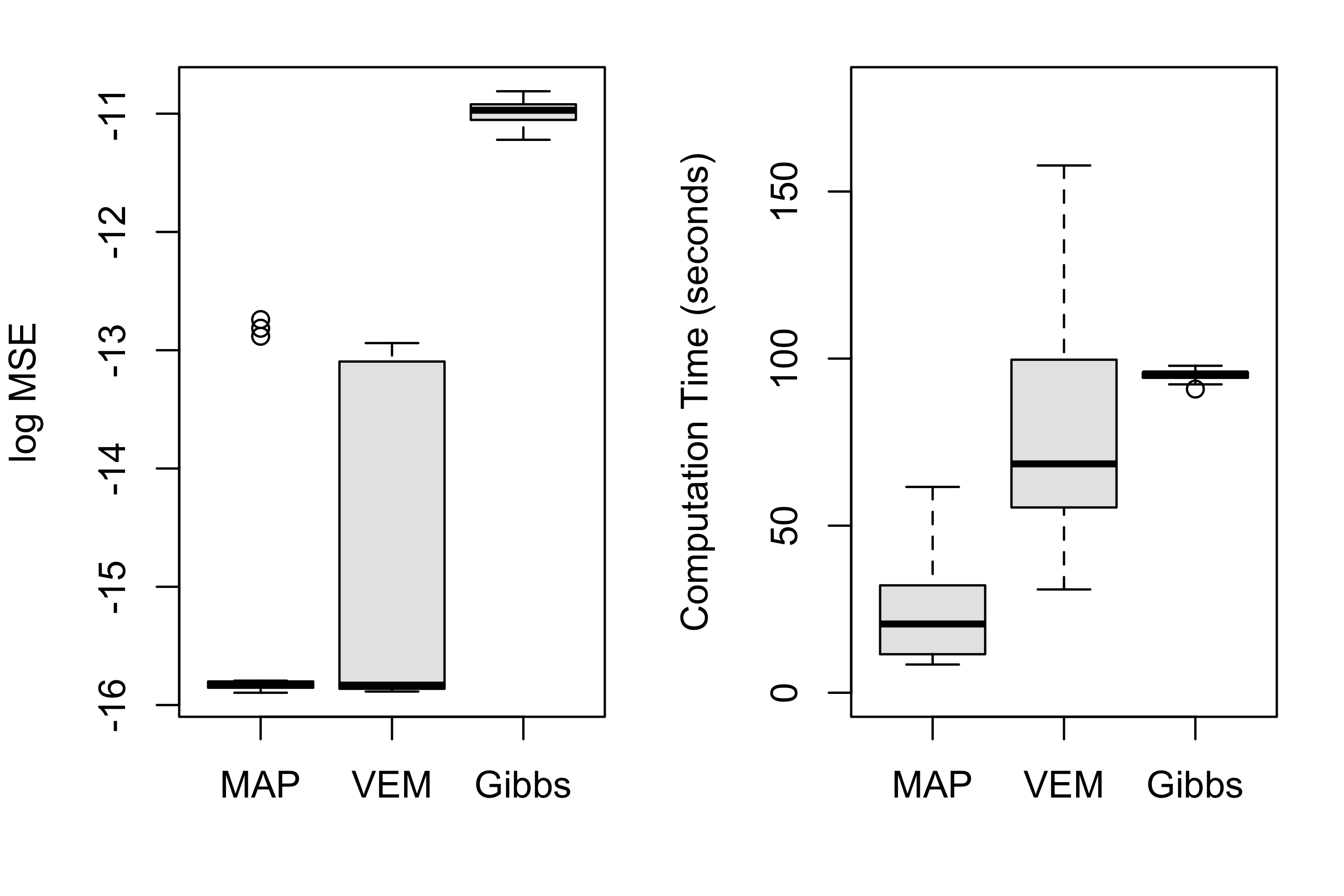}\\
\vskip -.5cm
\caption{\label{comparesim} Topic estimation given $K=10$ via MAP,
  VEM, and Gibbs procedures for data with $M=200$.  In (a), elements of
  each $\bs{\Theta}$ used in simulation are graphed against the
  corresponding estimates, and (b) shows the distributions for log
  MSE of these estimates and for algorithm computation time. }
\end{figure*}

\begin{figure*}[t!]
\vskip -.2cm
\hskip 2cm\includegraphics[height=2.1in]{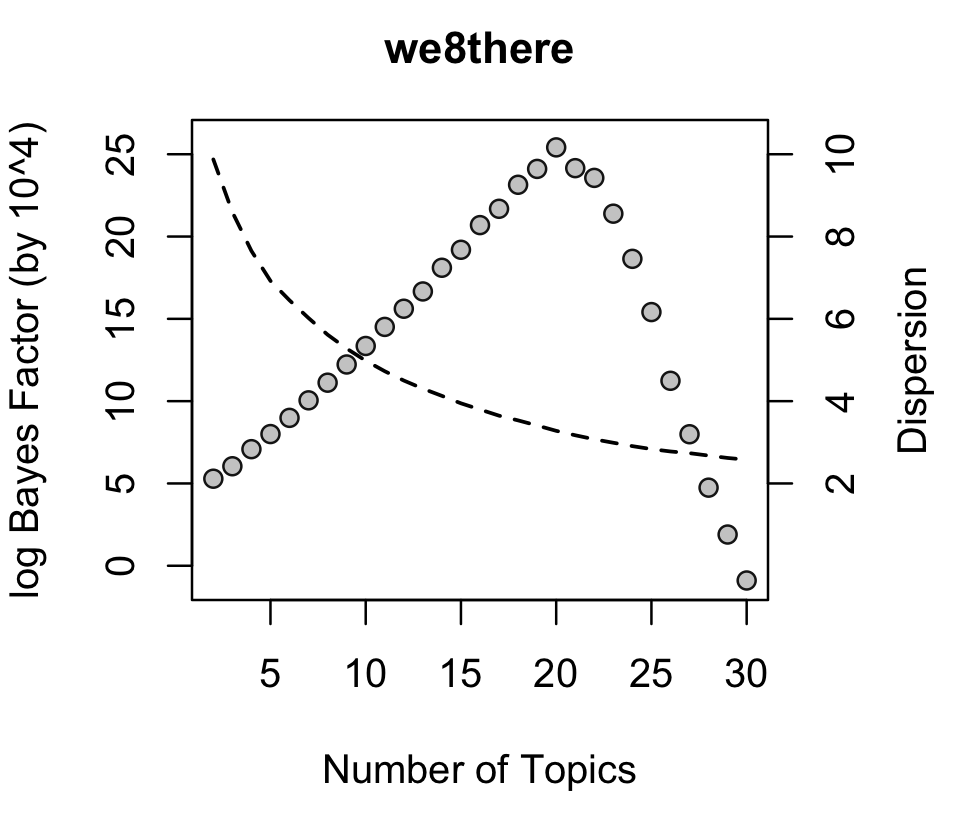}
\hskip 1.5cm\includegraphics[height=2.1in]{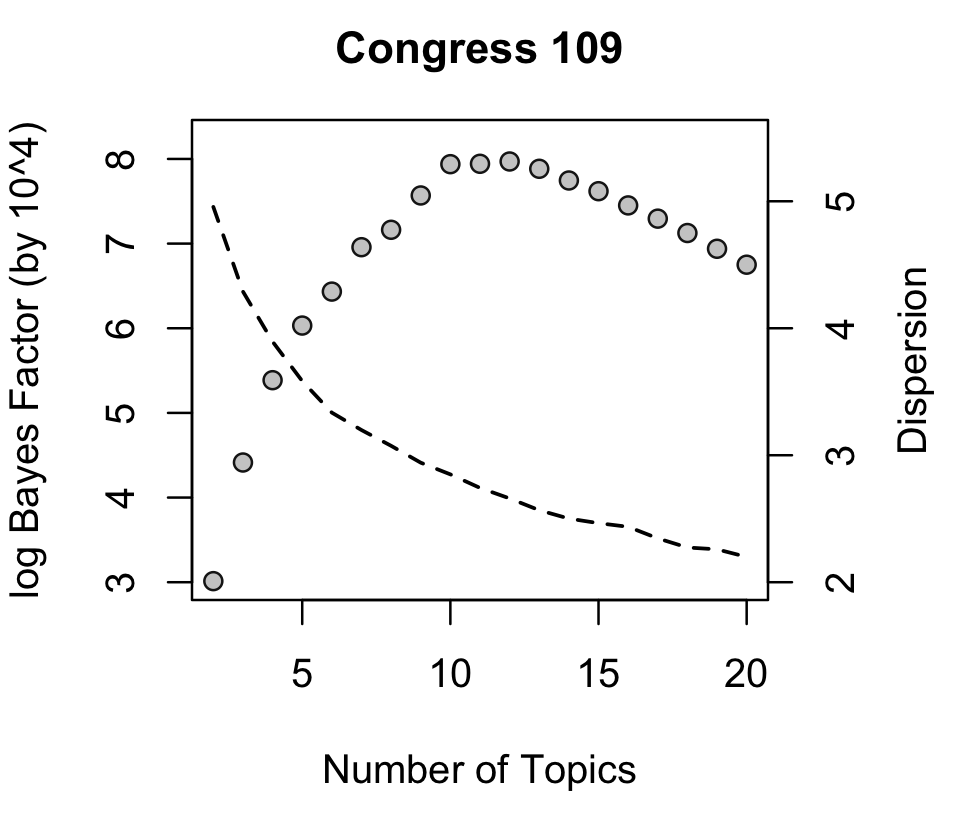}\\
\vskip -.8cm
\caption{\label{dataK} Model selection metrics for each dataset over a range of
  possible $K$.  In each case, the plotted points are log Bayes factor
  in multiples of $10^{4}$ and the dashed line is estimated
  dispersion.}
\vskip .3cm
\hskip 0.5cm\includegraphics[height=2.2in]{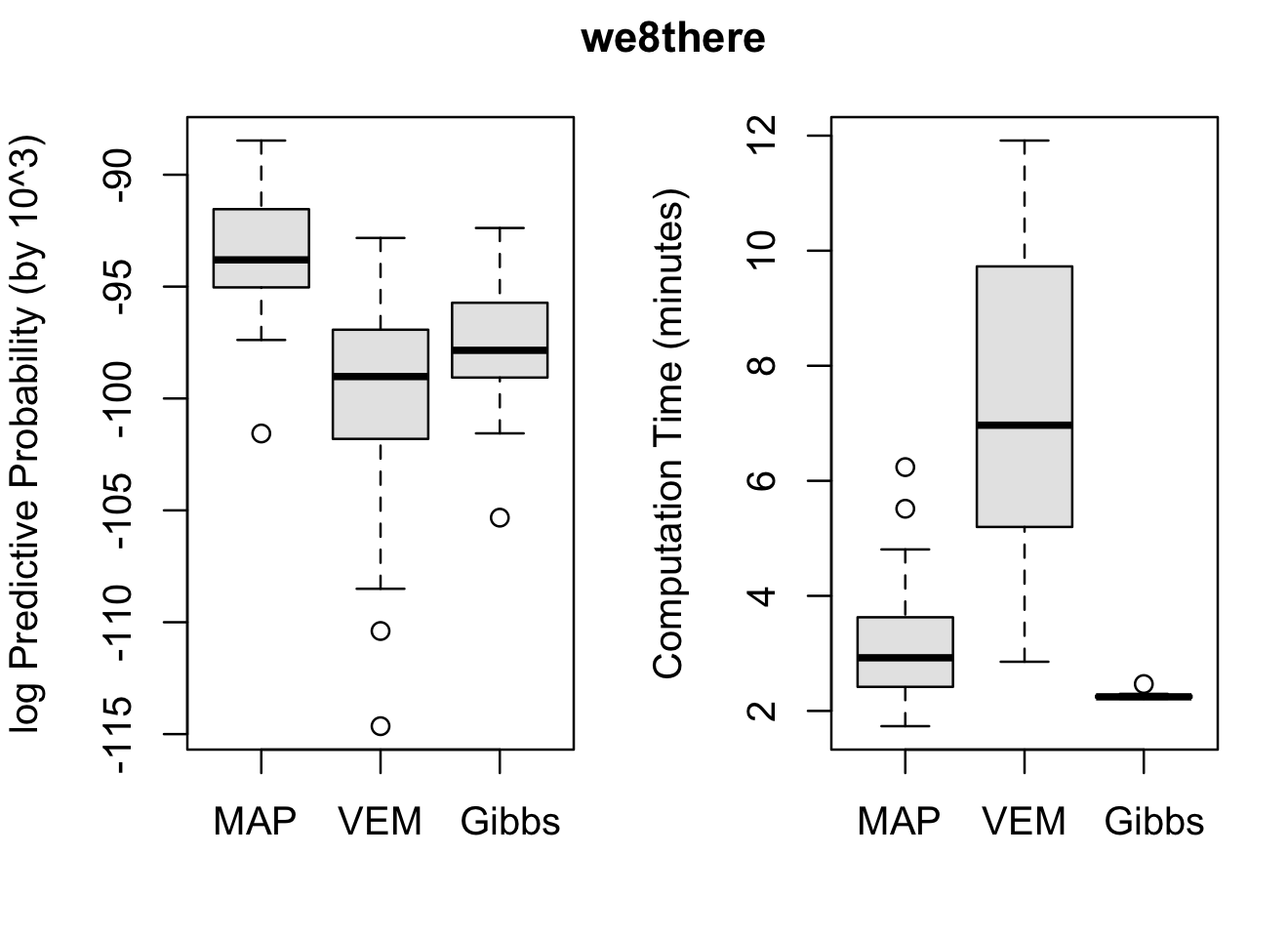}
\hskip 1cm\includegraphics[height=2.2in]{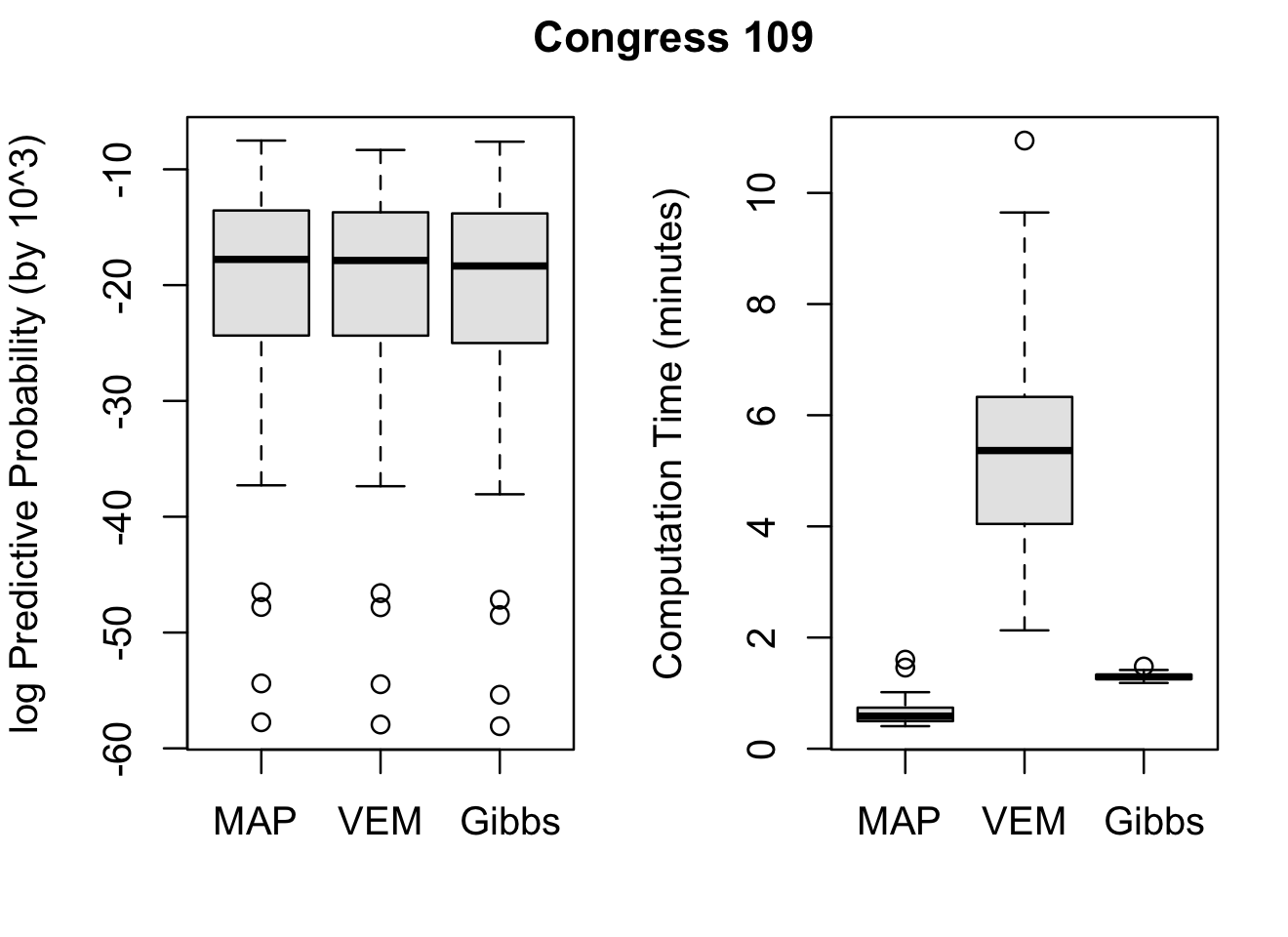}\\
\vskip -.6cm
\caption{\label{compareData} Out-of-sample performance for 50
  repetitions training on 80\% of data and validating on the left-out
  20\%, with predictive $\sum_{ij}x_{ij}\log(\bs{\theta}'_{\bullet
    j}\bs{\omega}_i)$ in multiples of $10^3$ and computation time in
  minutes.  }
\end{figure*}

\begin{figure*}[t!]
\hskip .5cm 
\fbox{
\begin{minipage}{6.2in}
\includegraphics[width=6.3in]{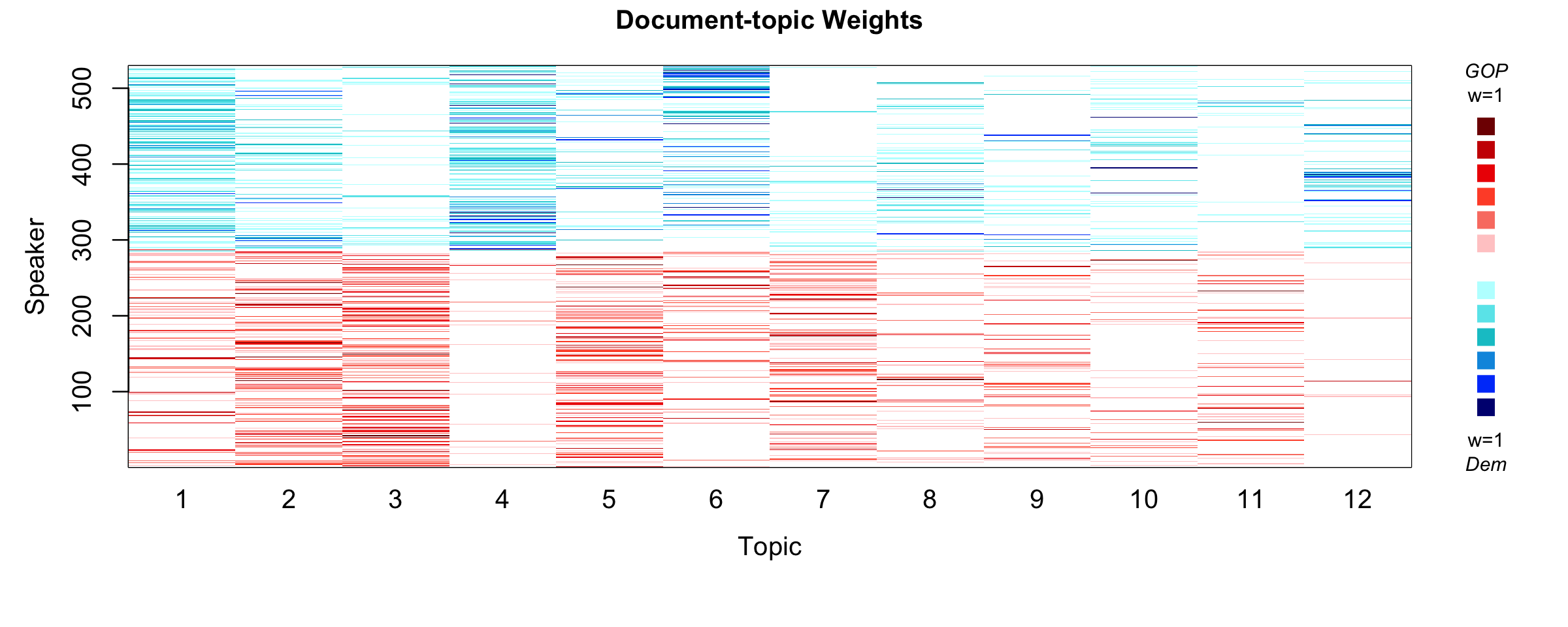}
\vspace{-.7cm}
\scriptsize \sf
\begin{enumerate}\setlength{\itemsep}{-.1cm} 
\item dropout.prevention.program, american.force.radio,
  national.endowment.art, head.start, flood.insurance.program (0.12)
\item near.earth.object, republic.cypru, winning.war.iraq, bless.america, troop.bring.home (0.12)
\item near.retirement.age, commonly.prescribed.drug, repeal.death.tax, increase.taxe, medic.liability.crisi (0.12)
\item va.health.care, united.airline.employe,
  security.private.account, private.account, issue.facing.american (0.11)
\item southeast.texa, temporary.worker.program, guest.worker.program,
  million.illegal.immigrant, guest.worker (0.11)
\item national.heritage.corridor, asian.pacific.american,
  columbia.river.gorge, american.heritage.month  (0.10)
\item ready.mixed.concrete, driver.education, witness.testify, indian.affair, president.announce (0.08)
\item low.cost.reliable, wild.bird, suppli.natural.ga, arctic.wildlife.refuge, price.natural.ga (0.06)
\item judicial.confirmation.process, fifth.circuit.court, chief.justice.rehnquist, summa.cum.laude, chief.justice (0.05)
\item north.american.fre, american.fre.trade, change.heart.mind,
  financial.accounting.standard, central.american.fre (0.05)
\item  pluripotent.stem.cel, national.ad.campaign, cel.stem.cel, produce.stem.cel, embryonic.stem (0.04)
\item able.buy.gun, deep.sea.coral, buy.gun,
  credit.card.industry, caliber.sniper.rifle  (0.04)
\end{enumerate}
\vskip .2cm
\end{minipage} }
\caption{\label{topics} Congress109 topics,
  summarized as their top-five terms by {\it lift} --
  $\theta_{kj}$ over empirical term probability -- and ordered
  by usage proportion (column-means of $\bs{\Omega}$, which are included
  in parentheses).  The image plot has cells shaded by magnitude
  of $\omega_{ik}$, with Republicans in red and Democrats in blue.  }
\end{figure*}

\section{Examples}
\label{examples}

Our methods are all implemented in the {\cd textir} package for {\cd
  R}.  For comparison, we also consider {\cd R}-package
implementations of VEM \citep[{\cd topicmodels
  0.1-1},][]{GrunHorn2011} and collapsed Gibbs sampling \citep[{\cd
  lda 1.3.1},][]{Chan2011}.  Although in each case efficiency could be
improved through techniques such as parallel processing or
thresholding for sparsity 
\citep[e.g., see the SparseLDA of][]{YaoMimnMcCa2009}, we seek to present a baseline comparison of
basic algorithms.  The priors of Section \ref{estimation} are used
throughout, and times are reported for computation on a 3.2 GHz Mac
Pro.

The original {\cd topicmodels} code measures convergence on {\it
  proportional} change in the log posterior bound, leading to observed
{\it absolute} log posterior change of 50-100 at termination under
tolerance of $10^{-4}$.  Such premature convergence leads to
very poor results, and the code ({\cd rlda.c} line 412) was altered to
match {\cd textir} in tracking absolute change.  Both routines then
stop on a tolerance of 0.1. Gibbs samplers were run 5000 iterations,
and {\cd lda} provides a single posterior draw for estimation.

\subsection{Simulation Study}

We consider data simulated from a ten-topic model with $p=1000$
dimensional topics $\bs{\theta}_k \sim \mr{Dir}(1/10),
~k=1,\ldots,10$, where each of $n=500$ documents are generated with
weights $ \bs{\omega}_i \sim \mr{Dir}(1/10)$ and phrase counts
$\bm{x}_i \sim \mr{MN}(\bs{\Theta}\bs{\omega}_i, m_i)$ given $m_i \sim
\mr{Po}(M)$.  This yields topics and topic weights that are dominated
by a subset of relatively large probability, and we have set sample
size at half vocabulary dimension to reflect the common $p>\!\!> n$
setting of text analysis.  Expected size, $M$, varies to illustrate
the effect of increased information in 5 sets of 50 runs for $M$ from
100 to 1600.

To investigate model selection, we fit $K= 5\ldots15$ topics to each
simulated corpus.  Figure \ref{simtopics} graphs results.  Marginal
likelihood estimates are in \ref{simtopics}.a as log Bayes factors,
calculated over the null model of $K=1$.  Apart from the low
information $M=100$, mean $\mr{p}(\bm{X} \mid K)$ is
maximized at the true model of $K=10$; this was very consistant across
individual simulations, with $K=10$ chosen invariably for $M\geq 200$.
Even at $M=100$, where $K=8$ was the most commonly chosen model, the
likelihood is relatively flat before quickly dropping for $K >
10$.  In \ref{simtopics}.b, the sample dispersion distribution is
shown for $K = 9,10,11$ at each size specification. Of primary
interest, $\hat\sigma^2$ is almost always larger than one for $K<10$
and less than one for $K\geq10$.  This pattern persists for un-plotted
$K$, and separation across models increases with $M$.  Estimated
dispersion does appear to be biased low by roughly 1-6\% depending on
size, illustrating the difficulty of choosing effective degrees of
freedom.  As a result, our $\chi^2$ test of a more-topics-than-$K$
alternative leads to $p$-values of $p=0$ for $K<10$ and
$p=1$ for $K\geq 10$.

We then consider MAP, VEM, and Gibbs estimation for the true $K=10$ model with
50 datasets simulated at $M=200$.  To account for label-switching,
estimated topics were matched with true topics by pairing the
$\bs{\Theta}$ columns of least sum squared difference.  Results are
shown in Figure \ref{comparesim}.  The plots in
\ref{comparesim}.a show largely accurate estimation for both VEM and
MAP procedures, but with near-zero $\theta$ occasionally fit at large
$\hat \theta$ and vice versa.  This occurs when some vocabulary
probabilities are swapped in estimation across topics, an event that
is not surprising under the weak identification of our
high-dimensional latent variable model.  It appears in
\ref{comparesim}.b that MAP does have slightly lower MSE, even though VEM
takes at least 2-3 times longer to converge.  Finally, as would be
expected of estimates based on a single draw from a short
MCMC run, Gibbs MSE are far larger than for either alternative.

\subsection{Data Analysis}

The data sets we consider are 
detailed in \citet{Tadd2012} and 
included as examples in {\cd textir}.  {\it
  We8there} consists of counts for 2804 bigrams in 6175
online restaurant reviews, and {\it
  Congress109} was compiled by \citet{GentShap2010} from the
$109^{th}$ US Congress as 529 legislators' usage counts for
each of 1000 bigrams and trigrams pre-selected for
partisanship.  

Model selection results are in Figure \ref{dataK}.  The marginal
likelihood surfaces, again expressed as Bayes factors, are maximized
at $K=20$ for the we8there data and at $K=12$ for congress109.
Interestingly, dispersion estimates remain larger than one for these
chosen models, and we do not approach $\hat \sigma^2 =1$ even for $K$
up to $200$.  This indicates alternative sources of overdispersion
beyond topic-clustering, such as correlation between counts across
phrases in a given document.

Working with the likelihood maximizing topic models, we then compare
estimators by repeatedly fitting $\bs{\widehat \Theta}$ on a random
80\% of data and calculating predictive probability over the left-out
20\%.  In each case, new document phrase probabilities were calculated
as $\bs{\widehat \Theta}\bs{\omega}_i$ using the conditional MAP for
$\bs{\omega}_i$ under a $\mr{Dir}(1/K)$ prior.  Figure
\ref{compareData} presents results.  As in simulation, the MAP
algorithm appears to dominate VEM: predictive probability is higher
for MAP in the we8there example and near identical across methods for
congress109, while convergence under VEM takes many times longer.
Gibbs sampling is quickly able to find a neighborhood of decent
posterior probability, such that it performs well relative to VEM but
 worse than the MAP estimators.

Finally, to illustrate the analysis facilitated by these models,
Figure \ref{topics} offers a brief summary of the congress109 example.
Top phrases from each topic are presented after ranking by term-lift,
$\theta_{kj}/q_j$ where $q_j = \sum_{i=1}^n x_{ij}/\sum_{i=1}^n m_i$,
and the image plots party segregation across topics.  Although
language clustering is variously ideological, geographical, and event
driven, some topics appear strongly partisan (e.g., 3 for Republicans
and 4 for Democrats).  Top-lift terms in the we8there example show
similar variety, with some topics motivated by quality, value or
service:

{ \scriptsize \sf\vskip -.1cm
1. anoth.minut, flag.down, over.minut, wait.over, arriv.after
  (0.07)}

\vskip -.3cm
{ \scriptsize \sf 2. great.place, food.great, place.eat, well.worth, great.price (0.06)
}

\vskip -.1cm
while others are associated with specific styles of food:

{ \scriptsize \sf \vskip -.1cm
9. chicago.style, crust.pizza, thin.crust, pizza.place, deep.dish (0.05)}

\vskip -.3cm
{ \scriptsize \sf 11. mexican.food, mexican.restaur, authent.mexican, best.mexican
(0.05). 
}

\vskip -.1cm
In both examples, the model provides massive dimension reduction (from
1000 to 12 and from 2804 to 20) by replacing individual phrases with
topic weights.

\section{Discussion}
\label{discussion}

Results from Section 5 offer some general support for our methodology.
In model selection, the marginal likelihood approximation
appears to provide for efficient data-driven selection of the number
of latent topics.  Since a default approach to choosing $K$ has been
thus far absent from the literature, this should be of use in the
practice of topic modeling.  We note that the same block-diagonal
Laplace approximation can be applied as a basis for inference and fast
posterior interval calculations. Dispersion shows potential for
measuring goodness-of-fit, but its role is complicated by bias and
alternative sources of overdispersion.

We were pleased to find that the efficiency of joint MAP estimation
did not lead to lower quality fit, but rather uniformly met or
outperformed alternative estimates.  The simple algorithm of this
paper is also straightforward to scale for large-data analyses.  In
a crucial step, the independent updates for each $\bs{\omega}_i
|\bs{\Theta}$ can be processed in parallel; our experience is that
this allows fitting 20 or more topics to hundreds of thousands of documents and
tens of thousands of unique terms in less than ten minutes on 
a common desktop.

\small
\bibliographystyle{chicago} \bibliography{taddy}

\begin{thebibliography}{}

\bibitem[\protect\citeauthoryear{Airoldi, Erosheva, Fienberg, Joutard, Love,
  and Shringarpure}{Airoldi et~al.}{2010}]{AiroErosFienJout2010}
Airoldi, E.~M., E.~A. Erosheva, S.~E. Fienberg, C.~Joutard, T.~Love, and
  S.~Shringarpure (2010).
\newblock Reconceptualizing the classification of pnas articles.
\newblock {\em Proceedings of the National Academy of Sciences\/}~{\em 107},
  20899---20904.

\bibitem[\protect\citeauthoryear{Alexander, Novembre, and Lange}{Alexander
  et~al.}{2009}]{AlexNoveLang2009}
Alexander, D.~H., J.~Novembre, and K.~Lange (2009).
\newblock Fast model-based estimation of ancestry in unrelated individuals.
\newblock {\em Genome Research\/}~{\em 19}, 1655--1664.

\bibitem[\protect\citeauthoryear{Asuncion, Welling, Smyth, and Teh}{Asuncion
  et~al.}{2009}]{AsunWellSmytTeh2009}
Asuncion, A., M.~Welling, P.~Smyth, and Y.~W. Teh (2009).
\newblock On smoothing and inference for topic models.
\newblock In {\em Proceedings of the Twenty-Fifth Conference on Uncertainty in
  Artificial Intelligence}. UAI.

\bibitem[\protect\citeauthoryear{Berger, Liseo, and Wolpert}{Berger
  et~al.}{1999}]{BergLiseWolp1999}
Berger, J.~O., B.~Liseo, and R.~L. Wolpert (1999).
\newblock Integrated likelihood methods for eliminating nuissance parameters.
\newblock {\em Statistical Science\/}~{\em 14}, 1--28.

\bibitem[\protect\citeauthoryear{Blei and Lafferty}{Blei and
  Lafferty}{2006}]{BleiLaff2006}
Blei, D.~M. and J.~D. Lafferty (2006).
\newblock Dynamic topic models.
\newblock In {\em Proceedings of the 23rd International Conference on Machine
  Learning}. ICML.

\bibitem[\protect\citeauthoryear{Blei and Lafferty}{Blei and
  Lafferty}{2007}]{BleiLaff2007}
Blei, D.~M. and J.~D. Lafferty (2007).
\newblock A correlated topic model of {S}cience.
\newblock {\em The Annals of Applied Statistics\/}~{\em 1}, 17--35.

\bibitem[\protect\citeauthoryear{Blei and McAuliffe}{Blei and
  McAuliffe}{2010}]{BleiMcAu2010}
Blei, D.~M. and J.~D. McAuliffe (2010).
\newblock Supervised topic models.
\newblock {arXiv:1003.0783v1}.

\bibitem[\protect\citeauthoryear{Blei, Ng, and Jordan}{Blei
  et~al.}{2003}]{BleiNgJord2003}
Blei, D.~M., A.~Y. Ng, and M.~I. Jordan (2003).
\newblock Latent {D}irichlet allocation.
\newblock {\em Journal of Machine Learning Research\/}~{\em 3}, 993--1022.

\bibitem[\protect\citeauthoryear{Chang}{Chang}{2011}]{Chan2011}
Chang, J. (2011).
\newblock {\em lda: Collapsed Gibbs sampling methods for topic models.}
\newblock R package version 1.3.1.

\bibitem[\protect\citeauthoryear{Deerwester, Dumais, Furnas, Landauer, and
  Harshman}{Deerwester et~al.}{1990}]{DeerDumaFurnLandHars1990}
Deerwester, S., S.~T. Dumais, G.~W. Furnas, T.~K. Landauer, and R.~Harshman
  (1990).
\newblock Indexing by latent semantic analysis.
\newblock {\em Journal of the American Society for Information Science\/}~{\em
  41}, 391--407.

\bibitem[\protect\citeauthoryear{Gentzkow and Shapiro}{Gentzkow and
  Shapiro}{2010}]{GentShap2010}
Gentzkow, M. and J.~Shapiro (2010).
\newblock What drives media slant? {E}vidence from {U.S.} daily newspapers.
\newblock {\em Econometrica\/}~{\em 78}, 35--72.

\bibitem[\protect\citeauthoryear{Griffiths and Styvers}{Griffiths and
  Styvers}{2004}]{GrifStyv2004}
Griffiths, T.~L. and M.~Styvers (2004).
\newblock Finding scientific topics.
\newblock In {\em Proceedings of the National Acadamy of Sciences}, Volume 101,
  pp.\  5228--5235. PNAS.

\bibitem[\protect\citeauthoryear{Grimmer}{Grimmer}{2010}]{Grim2010}
Grimmer, J. (2010).
\newblock A {B}ayesian hierarchical topic model for political texts: Measuring
  expressed agendas in senate press releases.
\newblock {\em Political Analysis\/}~{\em 18}, 1--35.

\bibitem[\protect\citeauthoryear{Gr\"{u}n and Hornik}{Gr\"{u}n and
  Hornik}{2011}]{GrunHorn2011}
Gr\"{u}n, B. and K.~Hornik (2011).
\newblock {\sf \small topicmodels}: {A}n {R} package for fitting topic models.
\newblock {\em Journal of Statistical Software\/}~{\em 40}, 1--30.

\bibitem[\protect\citeauthoryear{Haberman}{Haberman}{1973}]{Habe1973}
Haberman, S.~J. (1973).
\newblock The analysis of residuals in cross-classified tables.
\newblock {\em Biometrics\/}~{\em 29}, 205--220.

\bibitem[\protect\citeauthoryear{Hastie, Tibshrani, and Friedman}{Hastie
  et~al.}{2009}]{HastTibsFrie2009}
Hastie, T., R.~Tibshrani, and J.~H. Friedman (2009).
\newblock {\em The Elements of Statistical Learning}.
\newblock Springer.

\bibitem[\protect\citeauthoryear{Hofmann}{Hofmann}{1999}]{Hofm1999}
Hofmann, T. (1999).
\newblock Probabilistic latent semantic indexing.
\newblock In {\em Proceedings of the Twenty-Second Annual International SIGIR
  Conference}.

\bibitem[\protect\citeauthoryear{Ipsen and Lee}{Ipsen and
  Lee}{2011}]{IpseLee2011}
Ipsen, I. and D.~Lee (2011).
\newblock Determinant approximations.
\newblock {\sf \small http://arxiv.org/abs/1105.0437v1}.

\bibitem[\protect\citeauthoryear{Kass and Raftery}{Kass and
  Raftery}{1995}]{KassRaft1995}
Kass, R.~E. and A.~E. Raftery (1995).
\newblock Bayes factors.
\newblock {\em Journal of the American Statistical Association\/}~{\em 90},
  773--795.

\bibitem[\protect\citeauthoryear{Lange}{Lange}{2010}]{Lang2010}
Lange, K. (2010).
\newblock {\em Numerical Analysis for Statisticians\/} (2nd ed.).
\newblock Springer.

\bibitem[\protect\citeauthoryear{Luenberger and Ye}{Luenberger and
  Ye}{2008}]{Luen2008}
Luenberger, D.~G. and Y.~Ye (2008).
\newblock {\em Linear and Nonlinear Programming\/} (3rd ed.).
\newblock Springer.

\bibitem[\protect\citeauthoryear{Mackay}{Mackay}{1998}]{Mack1998}
Mackay, D. (1998).
\newblock Choice of basis for laplace approximation.
\newblock {\em Machine Learning\/}~{\em 33}, 77--86.

\bibitem[\protect\citeauthoryear{Neal}{Neal}{2000}]{Neal2000}
Neal, R. (2000).
\newblock Markov chain sampling methods for {D}irichlet process mixture models.
\newblock {\em Journal of Computational and Graphical Statistics\/}~{\em 9},
  249--265.

\bibitem[\protect\citeauthoryear{Pritchard, Stephens, and Donnelly}{Pritchard
  et~al.}{2000}]{PritStepDonn2000}
Pritchard, J.~K., M.~Stephens, and P.~Donnelly (2000).
\newblock Inference of polulation structure using multilocus genotype data.
\newblock {\em Genetics\/}~{\em 155}, 945--959.

\bibitem[\protect\citeauthoryear{Roeder and Wasserman}{Roeder and
  Wasserman}{1997}]{RoedWass1997}
Roeder, K. and L.~Wasserman (1997).
\newblock Practical {B}ayesian density estimation using mixtures of normals.
\newblock {\em Journal of the American Statistical Association\/}~{\em 92},
  894--902.

\bibitem[\protect\citeauthoryear{Taddy}{Taddy}{2012}]{Tadd2012}
Taddy, M.~A. (2011\setbox0=\hbox{2012}).
\newblock Inverse regression for analysis of sentiment in text.
\newblock {\em Submitted\/}.
\newblock {\sf\small http://arxiv.org/abs/1012.2098}.

\bibitem[\protect\citeauthoryear{Tang, Coram, Wang, Zhu, and Risch}{Tang
  et~al.}{2006}]{TangCoraWangZhuRisc2006}
Tang, H., M.~Coram, P.~Wang, X.~Zhu, and N.~Risch (2006).
\newblock Reconstructing genetic ancestry blocks in admixed individuals.
\newblock {\em American Journal of Human Genetics\/}~{\em 79}, 1--12.

\bibitem[\protect\citeauthoryear{Teh, Jordan, Beal, and Blei}{Teh
  et~al.}{2006}]{TehJordBealBlei2006}
Teh, Y.~W., M.~I. Jordan, M.~J. Beal, and D.~M. Blei (2006).
\newblock Hierarchical {D}irichlet processes.
\newblock {\em Journal of the American Statistical Association\/}~{\em 101},
  1566--1581.

\bibitem[\protect\citeauthoryear{Teh, Newman, and Welling}{Teh
  et~al.}{2006}]{TehNewmWell2006}
Teh, Y.~W., D.~Newman, and M.~Welling (2006).
\newblock A collapsed variational {B}ayesian inference algorithm for latent
  {D}irichlet allocation.
\newblock In {\em Neural Information Processing Systems}, pp.\  1--8. NIPS.

\bibitem[\protect\citeauthoryear{Tierney and Kadane}{Tierney and
  Kadane}{1986}]{TierKada1986}
Tierney, L. and J.~B. Kadane (1986).
\newblock Accurate approximations for posterior moments and marginal densities.
\newblock {\em Journal of the American Statistical Association\/}~{\em 81},
  82--86.

\bibitem[\protect\citeauthoryear{Wainwright and Jordan}{Wainwright and
  Jordan}{2008}]{WainJord2008}
Wainwright, M.~J. and M.~I. Jordan (2008).
\newblock Graphical models, exponential families, and variational inference.
\newblock {\em Foundations and Trends in Machine Learning\/}~{\em 1}, 1--305.

\bibitem[\protect\citeauthoryear{Wallach, Mimno, and McCallum}{Wallach
  et~al.}{2009}]{WallMimnMcCa2009}
Wallach, H.~M., D.~Mimno, and A.~McCallum (2009).
\newblock Rethinking {LDA}: Why priors matter.
\newblock In {\em Neural Information Processing Systems}. NIPS.

\bibitem[\protect\citeauthoryear{Yao, Mimno, and McCallum}{Yao
  et~al.}{2009}]{YaoMimnMcCa2009}
Yao, L., D.~Mimno, and A.~McCallum (2009).
\newblock Efficient methods for topic model inference on streaming document
  collections.
\newblock In {\em 15th Conference on Knowledge Discovery and Data Mining}.

\end{thebibliography}

\end{document}